# INTERCONNECTION OF DEFECT ENTROPIES AND ENTHALPIES IN $BAF_2$ REVISITED

Efthimios S. Skordas

*Department of Solid State Physics, Faculty of Physics, University of Athens, Panepistimiopolis, 157 84 Zografos, Greece*
*eskordas@phys.uoa.gr*

**Abstract**

Here, we investigate the following key prediction of a thermodynamical model that interrelates the defect parameters with the bulk elastic and expansivity data: for various defect processes in a given matrix material, a proportionality exists between defect entropies and enthalpies. The investigation is focused on $BaF_2$ for which ab-initio calculations within density functional theory and the generalized-gradient approximation have been recently made as far as the formation and migration of intrinsic defects is concerned, as well as for the elastic constants. Four defect processes have been studied in $BaF_2$: Anion Frenkel formation, fluorine vacancy migration, fluorine interstitial motion and electrical relaxation associated with a single tetravalent uranium. For these processes, the entropies and enthalpies vary by almost two orders of magnitude and reveal a proportionality between them. We find that this proportionality is solely governed by the bulk elasticity and expansivity data, which conforms to the aforementioned thermodynamical model.

## 1. Introduction

Despite their simple structure, alkaline-earth fluorides $XF_2$(X=Mg, Ca, Sr, Ba) with the cubic fluorite structure constitute an important class of relatively simple ionic crystals with a wide range of applications. As typical examples, $CaF_2$ and $BaF_2$, are wide band-gap materials of interest for their applications in precision vacuum ultraviolet lithography (Ref. 1 and references therein). A multitude of experimental and theoretical studies on their defect properties have shown that the dominant point defects are anion Frenkel pairs and ionic transport occurs mainly through migration of anion vacancies and interstitials[2,3].

Hereafter, we focus on $BaF_2$. It is presently known as one of the fastest inorganic scintillators for the detection of X-rays, gamma rays or other high energy particles with high efficiency[4]. One well known application is the detection of 511 KeV gamma photons in positron emission tomography[5]. $BaF_2$ has also been found to exhibit superionic conductivity, thus it has been considered as candidate material for high temperature batteries, fuel cells, chemical filters and sensors[6,7]. In addition, it has been found of usefulness in short wavelength lithography[8]. At ambient conditions $BaF_2$ crystallizes in the cubic phase. At pressures of about P=3-5 GPa it undergoes a phase transition to the orthorhombic phase[6, 9-11] and at about 10-15 GPa to the hexagonal phase[6, 10, 12].

In an earlier paper[13], we showed that the defect entropies and enthalpies in cubic BaF$_2$ for the anion Frenkel formation, the fluorine vacancy motion and the fluorine interstitial migration obtained from ionic conductivity measurements[14] (that have been carried out as in Refs. 15, 16) along with the defect migration parameters resulted from the analysis of the dielectric relaxation measurements[17] in BaF$_2$ doped with uranium, are interconnected in a way predicted by a model that interconnects the defect formation and migration parameters with bulk plasticity and expansivity data[18, 19]. In the meantime, new first principles density functional calculations in cubic phase in BaF$_2$ appeared within the generalized gradient approximation. These calculations during the last few years led to the determination of several properties including the elastic constants[20] as well as the formation and migration energies of intrinsic defects[21]. In the light of these new results, it was considered worthwhile to investigate again the extent of the validity of the aforementioned interconnection of the defect formation and migration parameters with the bulk elasticity and expansivity data.

The experimental values of the defect parameters along with the formation and migration enthalpies calculated within the frame of density-functional theory are compiled in Table 1 and plotted in Fig. 1.

## 2. The thermodynamical model and its comparison with experimental data revisited

The defect Gibbs energy $g^i$ is interconnected with the bulk properties of the solid through the relation:

$$g^i = c^i B \Omega \qquad (1)$$

where $B$ stands for the isothermal bulk modulus, $\Omega$ the mean volume per atom and $c^i$ is a dimensionless constant. The superscript $i$ refers to the defect process under investigation, e.g. defect formation, defect migration, self-diffusion activation. This model has been successfully applied for various defect processes, to several categories of solids including metals[19, 22], fluorides[23], mixed alkali halides[24, 25], diamond[26], oxides[27], semiconductors[28] as well as to complex materials which under uniaxial stress emit electric signals before fracture[29] similar to those detected[30, 31] before earthquakes[32].

By differentiating Eq.(1) with respect to temperature T, we find the entropy $s^i$ $(=(dg^i/dT)_P)$ and then inserting it into the relation $h^i = g^i + Ts^i$ the corresponding enthalpy $h^i$ is obtained. These two thermodynamic defect parameters are found to have a ratio[18]:

$$\frac{s^i}{h^i} = -\frac{\beta B + \frac{dB}{dT}\big|_P}{B - T\beta B - T\frac{dB}{dT}\big|_P} \qquad (2)$$

where, $\beta$ stands for the thermal volume expansion coefficient. Equation (2) reveals that the ratio $s^i/h^i$ should be the same for various defect processes in the same matrix material.

Here, we focus on the investigation of Eq.(2). The values of $B$ calculated at ambient conditions by different groups within the generalized gradient approximation during the last few years are compiled in Table 2. Their average value is ~56 GPa which differs slightly from the experimental $B$ value which is around 57 GPa (see Table 2). Thus, we adopt the latter value ($B \approx 57$ GPa) for the application of Eq.(2). By using also the experimental values[33], $\beta=0.55\times10^{-4}$K$^{-1}$, and $(d \ln B/dT)_P = -3.9\times10^{-4}$ K$^{-1}$, Eq. (2) leads to $s^i/h^i = 3.1\times10^{-4}$K$^{-1}$. This value corresponds to the straight line drawn in Fig.1.

An inspection of Fig. 1 shows that the experimental values related to two defect processes, i.e., the parameters corresponding to the dielectric relaxation as well as to those of the fluorine interstitial migration, are in reasonable agreement with Eq. (1) being almost on the straight line if the experimental uncertainty is taken into account. Concerning the other two defect processes, i.e., the fluorine vacancy motion and for the anion Frenkel formation, the points of which are marked with asterisk and open triangle respectively in Fig. 1, they seem to deviate somewhat from the straight line. For these two defect processes, we also insert in Fig. 1 the corresponding values (marked with a red inverted triangle and red diamond, respectively) –which are 0.59 eV and 1.88 eV respectively (see Table 1)- calculated within the density-functional theory by Nyawere et al.[21]. Interestingly, they both agree with the experimental values and scatter around the straight line predicted by Eq. (1).

## 3. Conclusion

Here, we studied the following four defect processes in BaF$_2$ for which experimental data were available: anion Frenkel formation, fluorine vacancy migration, fluorine interstitial motion and dielectric relaxation associated with a single tetravalent uranium. A proportionality emerges between the values of their defect entropies and defect enthalpies, which remarkably vary by two orders of magnitude. Moreover, it is found that this proportionality is governed by the bulk elastic and expansivity data as predicted by the thermodynamical cB$\Omega$ model. Our findings are strengthened by recent first principles calculations within the density functional theory and the generalized-gradient approximation.

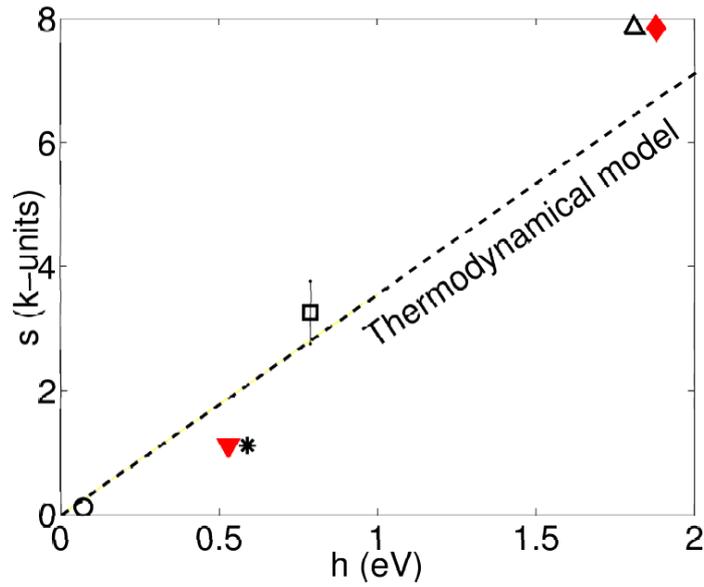

**FIGURE CAPTION**

The experimental values of the defect entropy s versus the defect enthalpy h for four defect processes in $BaF_2$. Open triangle: anion Frenkel formation; Open square: fluorine interstitial migration; Asterisk: fluorine vacancy motion; Open circle: electrical relaxation associated with a single tetravalent uranium. The solid red diamond and the solid red inverted triangle correspond to the calculated values within the density-functional theory for the enthalpy of the anion Frenkel formation and fluorine vacancy motion, respectively. The straight line results from the thermodynamic $cB\Omega$ model discussed in the text.

Table 1. Defect entropies and enthalpies in BaF$_2$

| Process | h (eV) | s (k$_B$-units) |
|---|---|---|
| Fluorine vacancy motion | 0.59[a], 053[b] | 1.12[a] |
| Fluorine interstitial migration | 0.79[a] | 3.26±0.49[a] |
| Anion-Frenkel formation | 1.81[a], 1.88[b] | 7.85[a] |
| Electrical relaxation associated with a single tetravalent uranium | 0.07[c] | 0.125[d] |

a. From the analysis of the ionic conductivity measurements[14]
b. Calculated value in Ref. 21
c. Experimental value from Ref. 17
d. By analyzing 13 the data published in Ref. 17

Table 2. Experimental (B$^{exp}$) and recently calculated (B$^{calc}$) values for the bulk modulus in BaF$_2$

| B$^{exp}$ GPa | B$^{calc}$ GPa |
|---|---|
| 57[10] | 53[a] |
| 56.9[31] | 60.6[b] |
| 57[34, 35] | 53.5[c] |

a. Calculated by Soni et al. (2011)[20]
b. Calculated by Fooladchang et al. (2013)[20]
c. Calculated by Nyawere et al. (2014)[20]